\documentclass[draft,onecolumn,12pt]{IEEEtran}

\usepackage{amsmath}   % From the American Mathematical Society

 \newtheorem{thm}{Theorem}%[subsection]
 \newtheorem{cor}{Corollary}
 \newtheorem{lem}{Lemma}
 \newtheorem{prop}{Proposition}

 \newtheorem{exam}{Example}
 \newenvironment{prf}{{\emph{Proof: }}}
 {\hfill QED.}

\begin{document}
\title{New Upper Bounds on Sizes of Permutation Arrays}
\author{Lizhen Yang, Ling Dong,
        Kefei Chen
    % <-this % stops a space
\thanks{Manuscript received June 14, 2006.
        This work was supported by NSFC under grants 90104005 and 60573030.}% <-this % stops a space
\thanks{Lizhen Yang is with the department of computer science and
engineering, Shanghai Jiaotong University, 800 DongChuan Road,
Shanghai, 200420, R.P. China (fax: 86-021-34204221, email:
lizhen\_yang@msn.com).}
\thanks{Ling Dong is with the department of computer science and
engineering, Shanghai Jiaotong University, 800 DongChuan Road,
Shanghai, 200420, R.P. China (fax: 86-021-34204221, email:
ldong@sh163c.sta.net.cn).}
\thanks{Kefei Chen is with the department of computer science and
engineering, Shanghai Jiaotong University, 800 DongChuan Road,
Shanghai, 200420, R.P. China (fax: 86-021-34204221, email:
Chen-kf@sjtu.edu.cn).}}
% note the % following the last \IEEEmembership and also the first \thanks -
% instance, "A\textbf{} \textbf{}B" will typeset as "A B" not "AB". If you want
% "AB" then you have to do: "A\textbf{}\textbf{}B"
\markboth{Journal of \LaTeX\ Class Files,~Vol.~1, No.~11,~November~2002}{Shell \MakeLowercase{\textit{et al.}}: Bare Demo of IEEEtran.cls for Journals}
% *** Note that you probably will NOT want to include the author's name in ***
% *** the headers of peer review papers.                                   ***

% If you want to put a publisher's ID mark on the page
% (can leave text blank if you just want to see how the
% text height on the first page will be reduced by IEEE)
%\pubid{0000--0000/00\$00.00~\copyright~2002 IEEE}

%\specialpapernotice{(Invited Paper)}

\maketitle

\begin{abstract}
A permutation array(or code) of length $n$ and distance $d$,
denoted by $(n,d)$ PA, is a set of permutations $C$ from some
fixed set of $n$ elements such that the Hamming distance between
distinct members $\mathbf{x},\mathbf{y}\in C$ is at least $d$. Let
$P(n,d)$ denote the maximum size of an $(n,d)$ PA. New upper
bounds on $P(n,d)$ are given. For constant $\alpha,\beta$
satisfying certain conditions, whenever $d=\beta n^{\alpha}$, the
new upper bounds are asymptotically better than the previous ones.
\end{abstract}

% Will be ignored for other modes.
\IEEEpeerreviewmaketitle

\section{Introduction}
\PARstart{L}{e}t $\Omega$ be an arbitrary nonempty infinite set.
Two distinct permutations $\mathbf{x},\mathbf{y}$ over $\Omega$
have distance $d$ if $\mathbf{x}\mathbf{y}^{-1}$ has exactly $d$
unfixed points.  A permutation array(permutation code, PA) of
length $n$ and distance $d$, denoted by $(n,d)$ PA, is a set of
permutations $C$ from some fixed set of $n$ elements such that the
distance between distinct members $\mathbf{x},\mathbf{y}\in C$ is
at least $d$. An $(n,d)$ PA of size $M$ is called an $(n,M,d)$ PA.
The maximum size of an $(n,d)$ PA is denoted as $P(n,d)$.

PAs are somewhat studies in the 1970s. A recent application by
Vinck ~\cite{Ferreira00,Vinck00Code,Vinck00Coded,Vinck00Coding} of
PAs to a coding/modulation scheme for communication over power
lines has created renewed interest in PAs. But there are still
many problems unsolved in PAs, e.g. one of the essential problem
is to compute the values of $P(n,d)$. It's known that determining
the exactly values of $P(n,d)$ is a difficult task, except for
special cases, it can be only to establish some lower bounds and
upper bounds on $P(n,d)$. In this correspondence, we give some new
upper bounds on $P(n,d)$,  which are asymptotically better than
the previous ones.

\subsection{Concepts and Notations}
We introduce concepts and notations that will be used throughout
the correspondence.

Since for two sets $\Omega,\Omega'$ of the same size, the
symmetric groups $Sym(\Omega)$ and $Sym(\Omega')$ formed by the
permutations over $\Omega$ and $\Omega'$ respectively, under
compositions of mappings, are isomorphic, we need only to consider
the PAs over $Z_n=\{0,1,\ldots,n-1\}$ and write $S_n$ to denote
the special group $Sym(Z_n)$. In the rest of the correspondence,
without special pointed out, we always assume that PAs are over
$Z_n$. We also write a permutation $\mathbf{a}\in S_n$ as an
$n-$tuple $(a_0,a_1,\ldots,a_{n-1})$, where $a_i$ is the image of
$i$ under $\mathbf{a}$ for each $i$. Especially, we write the
identical permutation $(0,1,\ldots,n-1)$ as $\mathbf{1}$ for
convenience. The Hamming distance $d(\mathbf{a},\mathbf{b})$
between two $n-$tuples $\mathbf{a}$ and $\mathbf{b}$ is the number
of positions where they differ. Then the distance between any two
permutations $\mathbf{x},\mathbf{y}\in S_n$ is  equivalent to
their Hamming distance.

Let $C$ be an $(n,d)$ PA. For an arbitrary permutation
$\mathbf{x}\in S_n$, $d(\mathbf{x},C)$ stands for the Hamming
distance between $\mathbf{x}$ and $C$, i.e.,
$d(\mathbf{x},C)=\min_{\mathbf{c}\in C}d(\mathbf{x},\mathbf{c})$.
A permutation in $C$ is also called a codeword of $C$. For
convenience for discussion, without loss of generality, we always
assume that $\mathbf{1}\in C$, and the indies of an $n-$tuple
(vector, array) are started by $0$. The support of a binary vector
$\mathbf{a}=(a_0,a_1,\ldots,a_{n-1})\in\{0,1\}^n$ is defined as
the set $\{i:a_i=1,i\in Z_n\}$, and the weight of $\mathbf{a}$ is
the size of its support, namely the number of ones in
$\mathbf{a}$. The support of a permutation
$\mathbf{x}=(x_0,x_1,\ldots,x_{n-1})\in S_n$ is defined as the set
of the points not fixed by $\mathbf{x}$, namely $\{i\in Z_n:
x_i\neq i\}$=$\{i\in Z_n: \mathbf{x}(i)\neq i\}$, and the weight
of $\mathbf{x}$, denoted as $wt(\mathbf{x})$, is defined as the
size of its support, namely the number of points in $Z_n$ not
fixed by $\mathbf{x}$.

A derangement of order $k$ is an element of $S_k$ with no fixed
points. Let $D_k$ be the number of derangements of order $k$, with
the convention that $D_0=1$. Then
$D_k=k!\sum_{i=0}^k\frac{(-1)^k}{k!}=\left[\frac{k!}{e}\right]$,
where $[x]$ is the nearest integer function, and  $e$ is the base
of the natural logarithm. The ball in $S_n$ of radius $r$ with
center $\mathbf{x}$ is the set of all permutations of distance
$\leq r$ from $\mathbf{x}$. The volume of such a ball is
\begin{equation}\label{eq:vol-PA}
V(n,r)=\sum_{i=0}^{r}{n\choose i}D_i.
\end{equation}

An $(n,d,w)$ constant-weight binary code is a set of binary
vectors of length $n$, such that each vector contains $w$ ones and
$n-w$ zeros, and any two vectors differ in at least $d$ positions.
The largest possible size of an $(n,d,w)$ constant-weight binary
code is denoted as $A(n,d,w)$. Similarly, we define an $(n,d,w)$
constant-weight PA as an $(n,d)$ PA such that each permutation is
of weight $w$, and denote the largest possible size of an
$(n,d,w)$ constant-weight PA as $P(n,d,w)$.

The concept of $P(n,d)$ can be further generalized. Let
$\Omega\subseteq S_n$, then $P_{\Omega}(n,d)$ denotes the maximum
size of an $(n,d)$ PA $C$ such that $C\subseteq\Omega$. For
trivial case $\Omega=S_n$, $P(n,d)=P_{\Omega}(n,d)$.

\subsection{Previous Results}
 The most basic upper bound on $P(n,d)$ is given by
Deza and Vanstone~\cite{Deza78}.
\begin{thm}~\cite{Deza78}.
\begin{equation}\label{upper:deza}
P(n,d)\leq \frac{n!}{(d-1)!}
\end{equation}
\end{thm}
We call the PAs which attain the Deza-Vanstone bound perfect PAs
and the known perfect PAs are
\begin{itemize}
    \item $(n,n,n)$ PAs for each $n\geq 1$;
    \item $(n,n!,2)$ PAs for each $n\geq 1$;
    \item $(n,n!/2,3)$ PAs for each $n\geq 1$~\cite{wensong04};
    \item $(q,q(q-1),q-1)$ PAs for each prime power
    $q$~\cite{Blake74};
    \item $(q+1,(q+1)q(q-1),q-1)$ PAs for each prime power
    $q$~\cite{Blake74};
    \item $(11,11\cdot10\cdot9\cdot8,8)$ PA~\cite{Blake74};
    \item $(12,12\cdot11\cdot10\cdot9\cdot8,8)$ PA~\cite{Blake74}.
\end{itemize}

The Deza-Vanstone bound can be derived by recursively applying the
following inequality.
\begin{prop}~\cite{wensong04}.\label{prop:n,n-1:prop:elementay:conse:P(n,d)}
\begin{equation}P(n,d)\leq
nP(n-1,d).\label{eq:n,n-1:prop:elementay:conse:P(n,d)}
\end{equation}
\end{prop}
Then for $d\leq m<n$, if we know $P(m,d)\leq M<\frac{m!}{(d-1)!}$,
we can get a stronger upper bound on $P(n,d)$:
\[
P(n,d)\leq \frac{n!P(m,d)}{m!}\leq\frac{n!M}{m!}.
\]
Another nontrivial upper bound on $P(n,d)$ is the sphere packing
bound obtained by considering the balls of radius $\lfloor
(d-1)/2\rfloor$~\cite{wensong04}.
\begin{thm}
\begin{equation}\label{eq:sphere-packing-upper-bound}
    P(n,d)\leq \frac{n!}{V(n,\lfloor (d-1)/2\rfloor)}.
\end{equation}
\end{thm}

For small values of $n$ and $d$, still stronger upper bounds are
founds in Tarnanen~\cite{Tarnanen99} by the method of linear
programming.

\subsection{Organization and New Results}
The correspondence is organized as follows. In Section II, we
first prove a relation between $P(n,d)$ and $P_{\Omega}(n,d)$ that
is the inequality
\[
P(n,d)\leq \frac{n!P_{\Omega}(n,d)}{|\Omega|}.
\]
Next, we give some elementary properties of $P(n,d,w)$, and then
use them to show a new upper bound on $P(n,d)$ for $d$ is even and
a new upper bound on $P(n,d)$ for $d$ is odd. They are given by
the following inequalities:
\[
P(n,2k)\leq \frac{n!}{V(n,k-1)+\frac{{n\choose k}D_k}{\lfloor
n/k\rfloor}}, \mbox{ for } 2\leq k\leq \lfloor n/2\rfloor;
\]
\begin{equation*}
P(n,2k+1)\leq \frac{n!}{V(n,k)+\frac{{n\choose
k+1}D_{k+1}-A(n-k,2k,k+1){n\choose k}D_k}{A(n,2k,k+1)}}, \mbox{
for } 2\leq k\leq \lfloor (n-k-1)/2\rfloor.
\end{equation*}
In Section III, we compare the upper bounds on $P(n,d)$ and show
for constant $\alpha,\beta$ satisfying certain conditions,
whenever $d=\beta n^{\alpha}$, the new upper bounds are
asymptotically better than the previous ones.
\section{The New Upper Bounds}
\begin{thm}\label{thm:Snboundsubset}
Let $\Omega$ be a subset of $S_n$. Then
\[
P(n,d)\leq \frac{n!P_{\Omega}(n,d)}{|\Omega|}.
\]
\end{thm}
\begin{prf}
Suppose $C$ is an $(n,P(n,d),d)$ PA. For any $\mathbf{x}\in S_n$,
let $\mathbf{x}C=\{\mathbf{x}\mathbf{c}:\mathbf{c}\in C\}$. Then
\begin{eqnarray*}
\sum_{\mathbf{x}\in S_n} |\mathbf{x}C\cap
\Omega|&=&\sum_{\mathbf{c}\in
C}\sum_{\mathbf{\omega\in\Omega}}|\{\mathbf{x}\in
S_n:\mathbf{x}\mathbf{c}=\omega\}|\\
&=&\sum_{\mathbf{c}\in C}\sum_{\mathbf{\omega\in\Omega}}|\{
\omega\mathbf{c}^{-1}\}|\\
&=&P(n,d)|\Omega|.
\end{eqnarray*}
On the other hand, there must exist $\mathbf{x}'\in S_n$ such that
$n!|\mathbf{x}'C\cap \Omega|\geq \sum_{\mathbf{x}\in
S_n}|\mathbf{x}C\cap \Omega|$. Then $n!|\mathbf{x}'C\cap
\Omega|\geq P(n,d)|\Omega|$, in other words, $P(n,d)\leq
\frac{n!|\mathbf{x}'C\cap \Omega|}{|\Omega|}$. This in conjunction
with $|\mathbf{x}'C\cap \Omega|\leq P_{\Omega}(n,d)$ results the
theorem.
\end{prf}

Since $S_{d}$ can be considered as a subset of $S_n$ for $d\leq
n$, Theorem~\ref{prop:n,n-1:prop:elementay:conse:P(n,d)} is also a
directly result of the above theorem, in fact
\[
P(n,d)\leq
\frac{|S_n|P(d,d)}{|S_d|}=\frac{n!d}{d!}=\frac{n!}{(d-1)!}.
\]
 The following is
also obtained immediately by Theorem~\ref{thm:Snboundsubset}.
\begin{cor}
\[
P(n,d)\leq \frac{n!P(n,d,w)}{{n\choose w}D_w}.
\]
\end{cor}

The following are well-known elementary properties of $A(n,d,w)$,
which will be applied to the proof of the properties of
$P(n,d,w)$.
\begin{lem}\label{lem:elem:proper:A(n,d,w)}
\begin{eqnarray*}
A(n,d,w)&=&1, \mbox{ if } d>2w;\\
A(n,2w,w)&=&\left\lfloor\frac{n}{w}\right\rfloor;\\
A(n,2k,k+1)&\leq& \left\lfloor \frac{n}{k+1}\left\lfloor
\frac{n-1}{k}\right\rfloor\right\rfloor.
\end{eqnarray*}
\end{lem}

\begin{thm}\label{thm:constant-weight-PA}
\begin{equation}\begin{array}{lll}
(I) &P(n,d,w)\leq A(n,2d-2w,w),&\mbox{for }d>w;\\
(II) &P(n,d,w)=1, &\mbox{for }d>2w,w\neq 1, d\geq 1;\\
(III) &P(n,2k,k)=\lfloor\frac{n}{k}\rfloor,&\mbox{for }2\leq
k\leq \lfloor n/2\rfloor;\\
(IV) &P(n,2k+1,k+1)=A(n,2k,k+1),&\mbox{for }1\leq
k\leq\lfloor (n-1)/2\rfloor;\\
(VI) &P(n,4,3)\leq \frac{2{n\choose 2}}{3},&\mbox{for }n\geq 4.\\
\end{array}
\end{equation}
\end{thm}
\begin{prf}
Part $(I)$ Let $C$ be an $(n,d,w)$ constant-weight PA with maximal
size $P(n,d,w)$, where $d>w$. Define $f:S_n\mapsto \{0,1\}^n$ such
that for any $\mathbf{a}=(a_0,a_1,\ldots,a_{n-1})\in S_n$ with
support $A$,
$f(\mathbf{a})=\mathbf{a'}=(a'_0,a'_1,\ldots,a'_{n-1})\in
\{0,1\}^n$, where
\begin{equation}\label{prf:thm:conweigt:eq1}
a'_i=\left\{\begin{array}{cl}
1,&\mbox{for }i\in A,\\
0,&\mbox{for }i\not\in A.
\end{array}
\right.
\end{equation}
Then $C'=\{f(\mathbf{a}): \mathbf{a}\in C\}$ is an $(n,2d-2w,w)$
constant-weight code with size $P(n,d,w)$ and this means
$P(n,d,w)\leq A(n,2d-2w,w)$. To prove this fact we need only to
prove that $C'$ have mutual distances $\geq 2d-2w$.

Let $\mathbf{a},\mathbf{b}\in C$, $\mathbf{a}\neq \mathbf{b}$, and
let $A$ and $B$ be the supports of $\mathbf{a}$ and $\mathbf{b}$
respectively. Suppose $\mathbf{a}'=f(\mathbf{a})$,
$\mathbf{b}'=f(\mathbf{b})$. (\ref{prf:thm:conweigt:eq1}) implies
\begin{eqnarray}
d(\mathbf{a}',\mathbf{b}')&=&|(A/B)\cup(B/A)|\nonumber\\
&=&|A|+|B|-2|A\cap B|\nonumber\\
&=&2w-2|A\cap B| \label{eq:prf:thm:weight2}
\end{eqnarray}
On the other hand, we have
\begin{equation*}
\begin{array}{lcl}
d&\leq& d(\mathbf{a},\mathbf{b})\\
&\leq& |A\cup B|\\
&=&|A|+|B|-|A\cap B|\\
&=&2w-|A\cap B|,
\end{array}
\end{equation*}
namely $|A\cap B|\leq 2w-d$. Putting this into
(\ref{eq:prf:thm:weight2}) we obtain
\[
d(\mathbf{a}',\mathbf{b}')\geq 2d-2w.
\]
Since $f$ is an onto mapping, we complete the proof of Part $(I)$.

Part $(II)$ For $d>2w,w\neq 1$ and $d\geq 1$, since
\begin{equation}\begin{array}{lcl}
2d-2w&>& 2\cdot 2w-2w=2w,
\end{array}
\end{equation}
$A(n,2d-2w,w)=1$ (by Lemma~\ref{lem:elem:proper:A(n,d,w)}) . This
in conjunction with part $(I)$ yields $P(n,d,w)=1$.

Part $(III)$ For $2\leq k\leq \lfloor n/2\rfloor$, by part $(I)$
 and Lemma~\ref{lem:elem:proper:A(n,d,w)} we have
\[
P(n,2k,k)\leq A(n,2k,k)=\lfloor n/k\rfloor.
\]

On the other hand, we can construct an $(n,2k,k)$ constant-weight
PA as follows:
\[
C=\{\mathbf{c}_i=(c_{i,0},c_{i,1},\ldots,c_{i,n-1})|i=0,1,\ldots,\lfloor
n/k\rfloor-1\},
\]
where
\[
c_{i,j}=\left\{\begin{array}{cl} j+1,& \mbox{for }
j=ik,ik+1,\ldots,ik+k-2\\
ik,&\mbox{for }j=ik+k-1\\
j,&\mbox{others}.
 \end{array}\right.
\]
Then we conclude $P(n,2k,k)=\lfloor n/k\rfloor$.

Part $(IV)$ For case $1\leq k\leq \lfloor (n-1)/2\rfloor$, by part
$(I)$ we have
\[
P(n,2k+1,k+1)\leq A(n,2k,k+1).
\]
Let $C'$ be an $(n,2k,k+1)$ constant-weight binary code with
maximal size $A(n,2k,k+1)$, then there exists $C\subseteq S_n$
such that for each member of $C$ there have one and only one
member of $C'$ with same support. We will  prove that $C$ is an
$(n,2k+1,k+1)$ constant-weight PA, which implies
$P(n,2k+1,k+1)\geq A(n,2k,k+1)$ and then results
$P(n,2k+1,k+1)=A(n,2k,k+1)$. Let $\mathbf{x},\mathbf{y}\in
C,\mathbf{x}\neq\mathbf{y}$ with corresponding supports $X$ and
$Y$. For case $X\cap Y=\O$,
$d(\mathbf{x},\mathbf{y})=|X|+|Y|=2k+2$. So we need only to
discuss the case $X\cap Y\neq \O$.  Let
$\mathbf{x}',\mathbf{y}'\in C'$ be the corresponding binary
codewords with supports $X,Y$. Since
$d(\mathbf{x}',\mathbf{y}')=|X|+|Y|-2|X\cap Y|=2(k+1)-2|X\cap
Y|\geq 2k$, $|X\cap Y|\leq 1$. Therefore, if $X\cap Y\neq \O$,
then $|X\cap Y|=1$. Suppose $X\cap Y=\{a\}$. Then
$\mathbf{x}(a)\neq \mathbf{y}(a)$, otherwise
$\mathbf{x}(a)=\mathbf{y}(a)=a$ and it lead to a contradiction.
Hence for this case, $d(\mathbf{x},\mathbf{y})=|A/B\cup
B/A|+|\{a\}|=|A/B|+|B/A|+1=2k+1$. Now we conclude that  $C$ is an
$(n,2k+1,k+1)$ constant-weight PA of size $A(n,2k,k+1)$, which
completes the proof of Part $(IV)$.

Part $(VI)$ Suppose $C$ is an $(n,4,3)$ constant-weight  PA. For
any pair $\{i,j\}\in Z_n\times Z_n$ with $i\neq j$, let
$C_{i,j}\subseteq C$ be the maximal set such that for each
$\mathbf{x}\in C_{i,j}$ with support $X$, $\{i,j\}\subseteq X$. We
are now ready to prove $|C_{i,j}|\leq 2$. Assume the contrary,
i.e., that $|C_{i,j}|\geq 3$ and
$\mathbf{x},\mathbf{y},\mathbf{z}$ are distinct elements of
$C_{i,j}$. W.l.o.g, $(\mathbf{x}(i),\mathbf{x}(j))=(k,i)$, where
$k\neq i,j$. Then $(\mathbf{y}(i),\mathbf{y}(j))=(j,k')$, where
$k'\neq i,j,k$, otherwise $d(\mathbf{x},\mathbf{y})<4$, which is a
contradiction. Similarly, $(\mathbf{z}(i),\mathbf{z}(j))=(j,k'')$,
where $k''\neq i,j,k,k'$. Thus $d(\mathbf{y},\mathbf{z})<4$, which
is a contradiction. Therefore $|C_{i,j}|\leq 2$.

Since there are ${n\choose 2}$ pairs of $(i,j)\in Z_n\times Z_n$
with $i\neq j$,
\begin{equation}\label{prf:thm:genboundweight5}
\sum_{i,j,i\neq j}|C_{i,j}|\leq 2{n\choose 2}.
\end{equation}
On the other hand, for each member of $C$, there are exactly 3
$C_{i,j}$ containing it, hence
\begin{equation}\label{prf:thm:genboundweight6}
\sum_{i,j,i\neq j}|C_{i,j}|=3|C|.
\end{equation}
Substituting (\ref{prf:thm:genboundweight6}) into
~(\ref{prf:thm:genboundweight5}) yields $|C|\leq \frac{2{n\choose
2}}{3}$, this means $P(n,4,3)\leq \frac{2{n\choose 2}}{3}$.
\end{prf}

\begin{thm}\label{thm:upper_bound_P(n,2k)}
For $2\leq k\leq \lfloor n/2\rfloor$,
\begin{equation}
P(n,2k)\leq \frac{n!}{V(n,k-1)+\frac{{n\choose k}D_k}{\lfloor
n/k\rfloor}}
\end{equation}
\end{thm}
\begin{prf} Let there be
$N_{k}$ permutations in $S_n$ which have distance $k$ to the
$(n,M,d)$ PA $C$. Then
\begin{equation}\label{eq:prf:thm:upper_bound_P(n,2k)}
MV(n,k-1)+N_k\leq n!
\end{equation}
In order to estimate $N_k$ we consider an arbitrary codeword
$\mathbf{c}$ which we can take to be $\mathbf{1}$(w.l.o.g.). Then
all permutations of weight $k$ has distance $k$ to $C$. Since
there are ${n\choose k}D_k$ permutations of weight $k$, there must
have ${n\choose k}D_k$ permutations that have distance $k$ to $C$.
By varying $\mathbf{c}$ we thus count $M{n\choose k}D_k$
permutations in $S_n$ that have distance $k$ to the PA. How often
has each of these permutations been counted. Take one of them;
again w.l.o.g. we call it $\mathbf{1}$. The codewords with
distance $k$ to $\mathbf{1}$ form an $(n,2k,k)$ constant-weight PA
since they have mutual distances $\geq 2k$ and weight $k$. Hence
there are at most $P(n,2k,k)=\lfloor n/k\rfloor$ (by part $(III)$
of Theorem~\ref{thm:constant-weight-PA}) such codewords. This
gives $N_k\geq \frac{M{n\choose k}D_k}{\lfloor n/k\rfloor}$.
Substituting this lower bounds on $N_k$ into
(\ref{eq:prf:thm:upper_bound_P(n,2k)}) implies the Theorem.
\end{prf}

\begin{thm}\label{thm:upper_bound_P(n,2k+1)}
For $2\leq k\leq \lfloor (n-k-1)/2\rfloor$,
\begin{equation}
P(n,2k+1)\leq \frac{n!}{V(n,k)+\frac{{n\choose
k+1}D_{k+1}-A(n-k,2k,k+1){n\choose k}D_k}{A(n,2k,k+1)}}
\end{equation}
\end{thm}
\begin{prf}Let $C$ be an $(n,M,2k+1)$ PA.
For any $\mathbf{x}\in S_n$, let
$B_i(\mathbf{x})=|\{\mathbf{c}:\mathbf{c}\in
C,d(\mathbf{c},\mathbf{x})=i\}|$.
 The proof relies on the following lemma.
\begin{lem}\label{lem:relation_for_prf_2k+1,upper}
\[
\begin{array}{ll}
A(n,2k,k+1)\sum_{i<k}B_{i}(\mathbf{x})+(A(n,2k,k+1)-A(n-k,2k,k+1))B_{k}(\mathbf{x})+B_{k+1}(\mathbf{x})\\
\leq A(n,2k,k+1)
\end{array}
\]
\end{lem}
\begin{prf}
Without loss of generality, we can take $\mathbf{x}=\mathbf{1}$,
then $B_i(\mathbf{x})$ is the number of codewords with weight $i$.
Clearly, a permutation with weight $w_1$ has distance $\leq
w_1+w_2$ to that with weight $w_2$. Hence
$\sum_{i<k}B_i(\mathbf{x})\leq 1$. If $B_i(\mathbf{x})>0$ for any
$i<k$, then $B_{k}(\mathbf{x})=B_{k+1}(\mathbf{x})=0$ and all the
other summands are zeros, and there is nothing to prove. Assume,
therefore, that $B_{i}(\mathbf{x})=0$ for all $i<k$. We know that
$B_k(\mathbf{x})\leq P(n,2k+1,k)=1$ (by part $(II)$ of
Theorem~\ref{thm:constant-weight-PA}), in other words
$B_k(\mathbf{x})$ is either 0 or 1: if it is 0, then the claim
becomes $B_{k+1}(\mathbf{x})\leq A(n,2k,k+1)=P(n,2k+1,k+1)$ (by
part $(IV)$ of Theorem~\ref{thm:constant-weight-PA}), which is
clear; if it is 1, then the claim becomes $B_{k+1}(\mathbf{x})\leq
A(n-k,2k,k+1)=P(n-k,2k+1,k+1)$, which is correct for there are no
points moved by both codewords of weight $k$ and of weight $k+1$.
\end{prf}

We are now ready to complete the proof of the theorem. It follows
from Lemma~\ref{lem:relation_for_prf_2k+1,upper} that
\begin{equation}\label{eq:prf:2k+1:upper1}
\begin{array}{l}
\sum_{\mathbf{x}\in
S_n}\left(A(n,2k,k+1)\sum_{i<k}B_{i}(\mathbf{x})+(A(n,2k,k+1)-A(n-k,2k,k+1))B_{k}(\mathbf{x})\right.\\
\left.+B_{k+1}(\mathbf{x})\right)\leq n!A(n,2k,k+1)
\end{array}
\end{equation}
The left side of the above inequality can be also written as
\begin{equation}\label{eq:prf:2k+1:upper2}
A(n,2k,k+1)\sum_{i<k}\sum_{\mathbf{x}\in
S_n}B_{i}(\mathbf{x})+(A(n,2k,k+1)-A(n-k,2k,k+1))\sum_{\mathbf{x}\in
S_n}B_{k}(\mathbf{x})+\sum_{\mathbf{x}\in S_n}B_{k+1}(\mathbf{x}).
\end{equation}
Now we shall give an expression in term of $M$ for
$\sum_{\mathbf{x}\in S_n}B_{i}(\mathbf{x})$. Since each codeword
$\mathbf{x}\in C$ has exactly ${n\choose i}D_i$ permutations in
$S_n$ which have distance $i$ to $\mathbf{x}$,  each codeword is
counted exactly ${n\choose i}D_i$ times by $\sum_{\mathbf{x}\in
S_n}B_{i}(\mathbf{x})$, which means
\begin{equation}\label{eq:prf:expression on M}
   \sum_{\mathbf{x}\in S_n}B_{i}(\mathbf{x})=M{n\choose i}D_i.
\end{equation}
Finally, the theorem is given by putting (\ref{eq:prf:expression
on M}) into ~(\ref{eq:prf:2k+1:upper2}), and rewriting
~(\ref{eq:prf:2k+1:upper1}) after replaced its left side
 by the new expression of (\ref{eq:prf:2k+1:upper2}).

\end{prf}

Using the upper bound on $A(n,2k,k+1)$ in
Lemma~\ref{lem:elem:proper:A(n,d,w)}, we get a determined upper
bound on $P(n,2k+1)$.
\begin{cor}\label{cor:upper_bound_MO_appx}
For $2\leq k\leq \lfloor(n-k-1)/2\rfloor$,
\[P(n,2k+1)\leq
\frac{n!}{V(n,k)+\frac{{n\choose k+1}D_{k+1}-\lfloor
\frac{n-k}{k+1}\lfloor\frac{n-k-1}{k}\rfloor\rfloor {n\choose
k}D_k}{\lfloor \frac{n}{k+1}\lfloor\frac{n-1}{k}\rfloor\rfloor}}
\]
\end{cor}

\section{Comparison of Upper Bounds}
In this section, we will prove that for constant $\alpha,\beta$
satisfying certain conditions, whenever $d=\beta n^{\alpha}$, the
new upper bounds on $P(n,d)$ are stronger than the previous ones
when $n$ large enough.

For large $n$ and $d$, the previous upper bounds on $P(n,d)$ have
Deza-Vanstone bound and sphere packing bound. Let $DV(n,d)$ denote
the Deza-Vanstone upper bound on $P(n,d)$ and $SP(n,d)$ denote the
sphere packing upper bound on $P(n,d)$, i.e.
\[DV(n,d)=\frac{n!}{(d-1)!},\]
\[
SP(n,d)=\frac{n!}{V(n,\lfloor (d-1)/2\rfloor)}.
\]
Although we can get more upper bounds on $P(n,d)$ by recursively
applying inequality (\ref{eq:n,n-1:prop:elementay:conse:P(n,d)})
and using the sphere packing bounds as the initial bound, namely,
for $d\leq m<n$,
\begin{equation}\label{eq:weaker:DV:SP}
P(n,d)\leq \frac{n!SP(m,d)}{m!},
\end{equation}
these bounds are not stronger than the best bounds given by
$DV(n,d)$ and $SP(n,d)$, which is shown as follows.
\begin{lem}For $d\leq m<n$,
\[
\frac{n!SP(m,d)}{m!}\geq \min\{DV(n,d),SP(n,d)\}.
\]
\end{lem}
\begin{prf}
If $SP(m,d)\geq DV(m,d)$,
$\frac{n!SP(m,d)}{m!}\geq\frac{n!}{m!}\cdot\frac{m!}{(d-1)!}=DV(n,d)$,
and there is nothing to prove. Therefore, assume
$SP(m,d)<DV(m,d)$. The claim is also correct since
\begin{eqnarray*}
  SP(n,d) &=& \frac{SP(n,d)}{SP(m,d)}\cdot SP(m,d) \\
   &=& \frac{n!}{m!}\cdot\frac{\sum_{i=0}^{\lfloor (d-1)/2\rfloor}{m\choose i}D_i}{\sum_{i=0}^{\lfloor (d-1)/2\rfloor}{n\choose i}D_i}\cdot SP(m,d) \\
   &<&\frac{n!SP(m,d)}{m!}.
\end{eqnarray*}
\end{prf}

Let $ME(n,k)$ denote the new upper bound on $P(n,2k)$ and
$MO(n,k)$ denote the new upper bound on $P(n,2k+1)$, i.e.
\[
ME(n,k)=\frac{n!}{V(n,k-1)+\frac{{n\choose k}D_k}{\lfloor
n/k\rfloor}},
\]
and
\[
MO(n,k)=\frac{n!}{V(n,k)+\frac{{n\choose
k+1}D_{k+1}-A(n-k,2k,k+1){n\choose k}D_k}{A(n,2k,k+1)}}.
\]

\begin{lem}\label{lem:compare:DV:PS}
For constants $\alpha,\beta$ satisfying either $0<\alpha<1/2$,
$\beta>0$ or $\alpha=1/2, 0<\beta<e$, whenever $d=\beta
n^{\alpha}$,
\[
\lim_{n\to\infty}\frac{DV(n,d)}{SP(n,d)}=\infty.
\]
\end{lem}
\begin{prf}
Let $k=\lfloor (d-1)/2\rfloor$. We have
\begin{eqnarray}
\lim_{n\to\infty}\frac{DV(n,d)}{SP(n,d)}&=&\lim_{n\to\infty}\frac{V(n,k)}{(d-1)!}\nonumber\\
&\geq&\lim_{n\to\infty}\frac{{n\choose
k}D_k}{(d-1)!}\nonumber\\
&=&\lim_{n\to\infty}\frac{n!k!}{ek!(n-k)!(d-1)!}\nonumber\\
&=&\lim_{n\to\infty}\frac{\sqrt{2\pi
n}(n/e)^n}{e\sqrt{2\pi(n-k)}((n-k)/e)^{n-k}\sqrt{2\pi
(d-1)}((d-1)/e)^{d-1}}\label{eq:prf:compare:DV:SP:1}
\end{eqnarray}
where the last equation is followed by Stirling's formula
$\lim_{n\to\infty}\frac{n!}{\sqrt{2\pi n}(\frac{n}{e})^n}=1$. By
(\ref{eq:prf:compare:DV:SP:1}),
\begin{equation}\label{eq:prf:compare:DV:SP:2}
\lim_{n\to\infty}\frac{DV(n,d)}{SP(n,d)}\geq\frac{1}{\sqrt{2\pi}}\lim_{n\to\infty}e^{d-k-2}\frac{n^{n+1/2}}{(n-k)^{n-k+1/2}(d-1)^{d-1/2}}.
\end{equation}
Let $c$ be a constant such that $c<1$. Since
\begin{eqnarray*}
  \lim_{n\to\infty}\left(\frac{n}{n-k}\right)^{\frac{n-k}{k}}&=&\lim_{n\to\infty}\left(1+\frac{1}{n/k-1}\right)^{n/k-1}\\
  &=&e,
\end{eqnarray*}
for $n$ large enough, $
\left(\frac{n}{n-k}\right)^{\frac{n-k}{k}}\geq e^c, $ i.e.
\begin{equation}\label{eq:forprove:DV:PS}
(n-k)^{n-k}\leq e^{-ck}n^{n-k}.
\end{equation}
Putting (\ref{eq:forprove:DV:PS}) into the right side of
(\ref{eq:prf:compare:DV:SP:2}), and multiplying the right side of
(\ref{eq:prf:compare:DV:SP:2}) by
$\lim_{n\to\infty}\frac{(n-k)^{1/2}}{n^{1/2}}=1$ and
\[\lim_{n\to\infty}\frac{(d-1)^{d-1/2}}{e^{-1}(\beta
n^{\alpha})^{d-1/2}}=\lim_{d\to\infty}e(\frac{d-1}{d})^{d-1/2}=\lim_{d\to\infty}e(1-1/d)^{d-1/2}=1,
\]
we obtain
\begin{eqnarray}
\lim_{n\to\infty}\frac{DV(n,d)}{SP(n,d)}&\geq&\frac{1}{\sqrt{2\pi}}\lim_{n\to\infty}e^{d-k-2}\frac{n^{n+1/2}}{e^{-ck}n^{n-k+1/2}e^{-1}(\beta n^{\alpha})^{(d-1/2)}}\nonumber\\
&=&\frac{1}{\sqrt{2\pi}}\lim_{n\to\infty}e^{d+(c-1)k-1}\beta^{-d+1/2}n^{k-\alpha
d+\alpha/2}\nonumber\\
&=&\frac{1}{\sqrt{2\pi}}\lim_{n\to\infty}e^{(1-\ln
\beta)d+(c-1)k-1+\ln\beta/2}n^{k-\alpha
d+\alpha/2}\nonumber\\
&\geq&\frac{1}{\sqrt{2\pi}}\lim_{n\to\infty}e^{d(\frac{1+c}{2}-\ln\beta)-1+\frac{\ln\beta}{2}
}n^{(1/2-\alpha)d-1+\alpha/2}\label{eq:prf:compare:DV:SP:3}
\end{eqnarray}
where the last inequality follows from $(c-1)k\geq (c-1)d/2$ and
\[k-\alpha
d+\alpha/2\geq (d/2-1)-\alpha
d+\alpha/2=(1/2-\alpha)d-1+\alpha/2.\]
To see the limit of right side of  (\ref{eq:prf:compare:DV:SP:3}), we discuss in two cases:\\
Case I:) $0<\alpha<1/2$. Since the coefficient $1/2-\alpha>0$, the
limit is determined by exponent $n^{(1/2-\alpha)d-1+\alpha/2}$,
and then the statement holds for this case.\\
Case II:) $\alpha=1/2,0<\beta<e$. The right side of
(\ref{eq:prf:compare:DV:SP:3}) is equal to
\[
\frac{1}{\sqrt{2\pi}}\lim_{n\to\infty}e^{d(\frac{1+c}{2}-\ln\beta)-1+\ln\beta/2}n^{-3/4}.
\]
The statement holds also, since for $0<\beta<e$ we can take $c$
such that $2\ln\beta-1<c<1$, in other words,
$$0<\frac{1+c}{2}-\ln\beta<1-\ln\beta.$$
\end{prf}
\begin{lem}\label{lem:compare:PS:ME}
For $k\geq 5$,
\[
SP(n,2k)-ME(n,k)>\frac{2(n-k+1)!}{n(k-1)}.
\]
\end{lem}
\begin{prf}Since $$V(n,k-1)+\frac{{n\choose k}D_k}{\lfloor n/k\rfloor}\leq
V(n,k-1)+{n\choose k}D_k=V(n,k),$$
\begin{eqnarray}
  SP(n,2k)-ME(n,k) &=& \frac{n!}{V(n,k-1)}-\frac{n!}{V(n,k-1)+\frac{{n\choose k}D_k}{\lfloor n/k\rfloor}}\nonumber \\
  &=&\frac{n!\cdot\frac{{n\choose k}D_k}{\lfloor n/k\rfloor}}{V(n,k-1)\left(V(n,k-1)+\frac{{n\choose k}D_k}{\lfloor
  n/k\rfloor}\right)}\nonumber\\
   &\geq&\frac{n!{n\choose k}D_k}{\lfloor n/k\rfloor V(n,k-1)V(n,k)}. \nonumber\label{eq:prf:compare:PS:ME:1}
\end{eqnarray}
When $k\geq 5$, $V(n,k-1)\leq (k-1){n\choose k-1}D_{k-1}$ and
$V(n,k)\leq k{n\choose k}D_k$, thereby
\begin{eqnarray}
  SP(n,2k)-ME(n,k) &\geq&\frac{n!{n\choose k}D_k}{\lfloor n/k\rfloor k{n\choose k}D_k(k-1){n\choose k-1}D_{k-1}} \nonumber\label{eq:prf:compare:PS:ME:2}\\
   &\geq&\frac{n!}{n(k-1){n\choose k-1}D_{k-1}}\label{eq:prf:compare:PS:ME:3}
\end{eqnarray}
When $k\geq 5$, $D_{k-1}=[(k-1)!/e]< \frac{(k-1)!}{2}$, putting
this into (\ref{eq:prf:compare:PS:ME:3}) we have
\begin{eqnarray*}
 SP(n,2k)-ME(n,k) &>&\frac{n!}{n(k-1)\cdot\frac{n!}{(n-k+1)!(k-1)!}\cdot\frac{(k-1)!}{2}}\\
   &=&\frac{2(n-k+1)!}{n(k-1)}.
\end{eqnarray*}
\end{prf}
\begin{thm}
For constants $\alpha,\beta$ satisfying either $0<\alpha<1/2$,
$\beta>0$ or $\alpha=1/2, 0<\beta<e$, whenever $2k=\beta
n^{\alpha}$,
\[
\lim_{n\to\infty}\left(\min\{DV(n,2k),SP(n,2k)\}-ME(n,k)\right)=\infty.
\]
\end{thm}
\begin{prf}By Lemma~\ref{lem:compare:PS:ME}, we have
\begin{eqnarray*}
  \lim_{n\to\infty} SP(n,2k)-ME(n,k)&\geq&\lim_{n\to\infty}\frac{2(n-k+1)!}{n(k-1)} \\
   &=&\lim_{n\to\infty}\frac{2(n-(\beta n^{\alpha})/2+1)!}{n((\beta n^{\alpha})/2-1)}  \\
   &=&\infty.
\end{eqnarray*}
By Lemma~\ref{lem:compare:DV:PS}, we have
\begin{eqnarray*}
  \lim_{n\to\infty}(DV(n,2k)-SP(n,2k)) &=&
  \lim_{n\to\infty}SP(n,2k)\left(\frac{DV(n,2k)}{SP(n,2k)}-1\right)\\
   &=& \infty,
\end{eqnarray*}
hence $\lim_{n\to\infty}(DV(n,2k)-ME(n,k))=\infty$, and then
follows the theorem.
\end{prf}

As a simple example of the superiority of the new bound $ME(n,k)$
over $DV(n,2k)$ and $SP(n,2k)$ we can compare them for small
values of $d$ and $n$.
\begin{exam}
$ME(20,4)<0.218\cdot 10^{15}$, $DV(20,8)>0.482\cdot 10^{15}$,
$SP(20,8)>0.984\cdot 10^{15}$, then $ME(20,4)$ provides the best
upper bound on $P(20,8)$.
\end{exam}
\begin{lem}\label{lem:compare:PS:MO}
For $k\geq 4$,
\[
SP(n,2k+1)-MO(n,k)>
\frac{2(n-k)!}{(k+1)n(n-1)}\left(1+k-\frac{n-1}{k}\right).
\]
\end{lem}
\begin{prf}We have
\begin{eqnarray}
  SP(n,2k+1)-MO(n,k) &=& \frac{n!}{V(n,k)}-\frac{n!}{V(n,k)+\frac{{n\choose
k+1}D_{k+1}-A(n-k,2k,k+1){n\choose k}D_k}{A(n,2k,k+1)}} \nonumber\\
   &=&\frac{n!\left(\frac{{n\choose
k+1}D_{k+1}-A(n-k,2k,k+1){n\choose
k}D_k}{A(n,2k,k+1)}\right)}{V(n,k)\left(V(n,k)+\frac{{n\choose
k+1}D_{k+1}-A(n-k,2k,k+1){n\choose k}D_k}{A(n,2k,k+1)}\right)} \nonumber\\
&\geq&\frac{n!\left(\frac{{n\choose k+1}D_{k+1}-
\frac{(n-k)(n-k-1)}{(k+1)k} {n\choose
k}D_k}{\frac{n(n-1)}{(k+1)k}}\right)}{V(n,k)V(n,k+1)}\label{eq:prf:compare:PS:MO:1}
\end{eqnarray}
where the last inequality is followed by $A(n-k,2k,k+1)\leq
\frac{(n-k)(n-k-1)}{(k+1)k}$,
$A(n,2k,k+1)\leq\frac{n(n-1)}{(k+1)k}$ (by
Lemma~\ref{lem:elem:proper:A(n,d,w)}) and
$$V(n,k)+\frac{{n\choose k+1}D_{k+1}-A(n-k,2k,k+1){n\choose
k}D_k}{A(n,2k,k+1)}\leq V(n,k)+{n\choose k+1}D_{k+1}=V(n,k+1).$$
When $k\geq 4$, $V(n,k)\leq k{n\choose k}D_k$  and $V(n,k+1)\leq
(k+1){n\choose k+1}D_{k+1}$, then
\begin{eqnarray*}
SP(n,2k+1)-MO(n,k) &\geq&\frac{n!\left(\frac{{n\choose
k+1}D_{k+1}-\frac{(n-k)(n-k-1)}{(k+1)k}{n\choose
k}D_k}{\frac{n(n-1)}{(k+1)k}}\right)}{k{n\choose
k}D_k(k+1){n\choose k+1}D_{k+1}}\\
&=&\frac{(n-2)!\left(\frac{D_{k+1}}{D_k}-\frac{n-k-1}{k}\right)}{{n\choose
k}D_{k+1}}
\end{eqnarray*}
Since for $k\geq 4$, $\frac{D_{k+1}}{D_k}\geq
\frac{(k+1)!/e-1}{k!/e+1}=k+1-\frac{k+2}{k!/e+1}> k$, and
$D_{k+1}\leq \frac{(k+1)!}{e}+1<(k+1)!/2$,
\begin{eqnarray*}
SP(n,2k+1)-MO(n,k)&>&\frac{(n-2)!\left(k-\frac{n-k-1}{k}\right)}{{n\choose
k}(k+1)!/2}\\
&=&\frac{2(n-k)!}{(k+1)n(n-1)}\left(1+k-\frac{n-1}{k}\right).
\end{eqnarray*}
\end{prf}

\begin{thm}
For constant $\beta$ such that $2<\beta<e$, whenever $2k+1=\beta
n^{1/2}$,
\[
\lim_{n\to\infty}\left(\min\{DV(n,2k+1),SP(n,2k+1)\}-MO(n,k)\right)=\infty.
\]
\end{thm}
\begin{prf}Since
\begin{eqnarray*}
  1+k-\frac{n-1}{k} &\geq&1+\frac{2\sqrt{n}-1}{2}-\frac{n-1}{\frac{2\sqrt{n}-1}{2}}\\
   &=&1+\left(\sqrt{n}-\frac{1}{2}\right)-\left(\sqrt{n}-\frac{1}{2}+\frac{\sqrt{n}-\frac{5}{4}}{\sqrt{n}-\frac{1}{2}}\right) \\
   &=&\frac{3}{4\sqrt{n}-2},
\end{eqnarray*}
by Lemma~\ref{lem:compare:PS:MO} we have
\begin{eqnarray}
  \lim_{n\to\infty}SP(n,2k+1)-MO(n,k)&\geq&\lim_{n\to\infty}
\frac{2(n-k)!}{(k+1)n(n-1)}\left(1+k-\frac{n-1}{k}\right)\nonumber\\
 &\geq&\lim_{n\to\infty}\frac{2(n-k)!}{(k+1)n(n-1)}\cdot\frac{3}{4\sqrt{n}-2}\label{eq:prf:compare:MO}\\
   &=&\infty.\nonumber
\end{eqnarray}

By Lemma~\ref{lem:compare:DV:PS}, we have
\begin{eqnarray*}
  \lim_{n\to\infty}(DV(n,2k+1)-SP(n,2k+1)) &=&
  \lim_{n\to\infty}SP(n,2k+1)\left(\frac{DV(n,2k+1)}{SP(n,2k+1)}-1\right)\\
   &=&\infty,
\end{eqnarray*}
hence $\lim_{n\to\infty}(DV(n,2k+1)-MO(n,k))=\infty$, and then
follows the theorem.
\end{prf}

As a simple example of the superiority of the new bound $MO(n,k)$
over $DV(n,2k+1)$ and $SP(n,2k+1)$ we can compare them for small
values of $d$ and $n$.
\begin{exam}
$MO(20,4)<0.380\cdot 10^{14}$ by
Corrollary~\ref{cor:upper_bound_MO_appx}, $SP(20,9)>0.528\cdot
10^{14}$, $DV(20,9)>0.603\cdot 10^{14}$, then $MO(20,4)$ provide
the best upper bound on $P(20,9)$.
\end{exam}

\hfill mds

\hfill November 18, 2002

\bibliographystyle{IEEEtran}
\bibliography{IEEEabrv,bib}
\end{document}